**Position statement for HCII Workshop: 8:30 am – 12:30 pm, Tuesday, 28 July 2026**
HCI Design in an AI-empowered World

# Control panels to clarify user intent with Large Language Models

Ben Shneiderman, April 30, 2026

**Abstract:** Typical user interfaces for Large Language Models present a blank prompt window that invites a natural language query by users, but offers little guidance. This paper proposes a visual control panel interface that would provide more cues to the semantics of prompt formation, enabling users to more easily express their intent. By emphasizing recognition over recall, control panels help users formulate more effective prompts that match their intent.

## Introduction

The essential tasks of information seeking research are to understand how people come to seek information, how they refine their understanding, express their intent, extend their knowledge, and then use the new knowledge to take action and possibly seek further information.

A widely held current belief is that Agentic AI will facilitate human epistemic exploration and resolve users' information needs by way of natural language dialogs thru Large Language Models (LLMs). Many people consider products, such as OpenAI's ChatGPT, Anthropic's Claude, Microsoft's Copilot, or Google's Gemini, as strong examples of how AI can respond to user needs. While agentic definitions vary from broad ones that include any process with an input and an output to other definitions that suggest that agents are like travel agents or butlers who respond to requests from people.

## Related work

Ryen White (2024) described the widely held belief that "Natural language is an expressive and powerful means of communicating intentions and preferences with search systems. … The ability of agents to better understand intentions and provide assistance beyond fact finding … will advance the search frontier." He argues that agents will relieve users of the need to decompose their high-level tasks into specific subtasks ("By decomposing tasks into subtasks and assigning them to specialized agents, we can manage complexity more effectively") and concludes confidently that "AI agents will transform how we search."

A follow up article, co-authored with Chirag Shah (2025), carried the theme further: "An agent … is an autonomous entity or program that takes preferences, instructions, or other forms of inputs from a user to accomplish specific tasks on that user's behalf." They review the half century of work on agents, identify the strengths and weaknesses of current agents, and propose a new ecosystem that will “represent the next age of evolution for capable AI systems… that go beyond information retrieval and generation to perform deep reasoning and take actions on a user’s behalf.” I encouraged attendees to read these two well-organized and well-written papers.

A few months later, Shah and Lynda Tamine (2025) wrote: “we argue that this metaphor [human-AI collaboration] fundamentally misrepresents these interactions and obscures critical issues of agency, accountability, and labor.”

At the ACM CHIIR Conference Shah (March 2026) positioned AI systems as educational scaffolds that support learning rather than bypass it. The paper ends well with "we must advocate for human information seeking as a valuable cognitive activity – not merely a means to acquire information but a process through which understanding emerges."

I wrote to Shah before the conference: “For me, humans have agency, goals, intentions, desires, and abilities, while AI, IR, and any computing machine is just a tool (or a supertool). The use of the term ‘agent’ has elevated computers to be like a human, but that is the source of the many misconceptions, which you describe. ... I think we are largely in agreement that people and computers are different, that only people have intentions, and that human learning is a desirable product of search. I would add that humans take pleasure from learning and are delighted by new-found understandings.”

Shah wrote back “Where I would gently push back: I think the ‘just a tool’ framing, while correct at the ontological level, may understate the practical stakes. A hammer does not reshape what users believe they know, alleviate their motivation to learn, or redistribute epistemic credit in ways that obscure accountability. These AI systems do all three — not because they have agency, but precisely because they are deployed as if they do. The ‘supertool’ framing you offer actually captures this well, and I wish I had used it.”

**My claims**

My keynote described this debate and argued that instead of natural language exchanges, users could express their intents by starting with a natural language request, accompanied by a control panel that reminded users of other possibilities. Intent refinement would be facilitated by revisions to the natural language statements and further selections from additional control panels, customized to users’ needs.

The inclusion of compact visual control panels that favor recognition over recall and present abundant choices facilitate understanding and forming mental models. Control panels could include buttons, pulldowns, check boxes, sliders, form fill-in, and other direct manipulation controls, which facilitate exploration by way of easily reversible actions and feedback that clarifies possibilities. AI technologies can be used to set defaults, adjust for context, and make recommendations, applying the design principles of Human-Centered AI (Shneiderman 2022).

Successful control panel interfaces include restaurant reservations systems (OpenTable, Resy), airlines (United, Air Canada, Delta), hotels (Hilton, Marriott, Holiday Inn), and shopping (Amazon). However, current designs for LLMs (ChatGPT, Copilot, Claude) rely on a simple blank prompt window that provides little guidance to users who may have an ill-formed intent.

One aspect of the hidden features is that there are 25 styles built in to OpenAI's ChatGPT, but most users are unaware of these hidden features, so they are unlikely to specify these possibilities. A visual control panel would give users greater control by emphasizing recognition over recall (Figure 1), provide some examples, and put the emphasis on the user.

What do you want?
Coming of age story about a teenage boy in Seattle in 1960
Job application for a project manager in social media
+ Ask for Anything
Goal
Fiction
Historical
Journalism
Resume
Ad
Poetry
Speech
See more
Style
Academic
Concise
Conversational
Technical
Humorous
Motivational
See more
Tone
Authoritative
Critical
Optimistic
Sarcastic
Mystical
Empathetic
Flowery
See more
Format
Paragraphs
Bullet list
Multi-column table
Limerick
Sonnet
Dialog
See more
Length: 10 – 1000 words

*Figure 1: Proposed interface for ChatGPT that exposes some of the goals, styles, tones, and formats that users might choose.*

This interface removes the suggestion that there is a human-like agent, giving users greater control, which supports their self-efficacy, enables creative exploration, and clarifies their responsibility for the results. I believe that this style of interface would result in more users taking on more ambitious tasks and experts becoming SuperExperts.

Shah wrote that "Perhaps the most useful next step is not another framework paper, but a design competition: who can build a search interface that is both genuinely useful and measurably better for learning than a blank prompt window?"

After the conference White wrote "you made an interesting point about people wanting information to perform action (the information is rarely an end in itself). … A key question is clearly how to keep people at the center of these AI advances: empowered, in control, learning, and able to benefit from what automation makes possible. I find the control/agency/engagement aspects especially interesting given AI's emerging abilities to tackle long-running, complex tasks and new mechanisms for coordinating agent work."

## Conclusion

Visual control panel interfaces help users clarify their intent and use the powerful features built into Large Language Models. These interfaces apply theories of information seeking and learning, as well as benefitting users.